# Revealing photon's behaviors in a birefringent interferometer

Zhi-Yuan Zhou,[1,2,3,†,*] Shi-Kai Liu,[1,2,†] Shi-Long Liu,[1,2] Yin-Hai Li,[1,2,3] Yan Li,[1,2] Chen Yang,[1,2] Zhao-Huai Xu,[1,2] Guang-Can Guo,[1,2] and Bao-Sen Shi[1,2,3,*]

[1]*CAS Key Laboratory of Quantum Information, USTC, Hefei, Anhui 230026, China*
[2]*Synergetic Innovation Center of Quantum Information & Quantum Physics, University of Science and Technology of China, Hefei, Anhui 230026, China*
[3]*Wang Da-Heng Collaborative Innovation Center for Science of Quantum Manipulation & Control, Heilongjiang Province & Harbin University of Science and Technology, Harbin 150080, China*

*Corresponding authors: zyzhouphy@ustc.edu.cn; drshi@ustc.edu.cn*

[†] *These two authors contributed equally to this work.*

The interferometer is one of the most important devices for revealing the nature of light and for precision optical metrology. Though lots of experiments were performed for probing photon's behaviors in various configurations, a complete study of photon's behavior in a birefringent interferometer has not ever been performed. Based on an environmental turbulence immune Mach-Zehnder interferometer, tunable photonic beatings by rotating a birefringent crystal versus the temperature of the crystal for both single-photon and two-photon are observed. Furthermore, the two-photon interference fringes beat two times faster than the single-photon interference fringes. This beating effect is used to determine the thermal dispersion coefficients of the two principal refractive axes with a single measurement, the two-photon interference shows super-resolution and high sensitivity. Obvious differences between two-photon and single photon interference are also revealed in an unbalanced situations. In addition, influences of the photon bandwidth to the beating behaviors that come from polarization-dependent decoherence are also investigated. Our findings will be important for better understanding the behavior of two-photon interference in a birefringent interferometer and for precision optical metrology with quantum enhancement.

Since Young's pioneer work of double slits interference in 1807 [1, 2], the optical interferometer is a basic tool of modern science and technology that can be used to study the fundamental nature of light and has broad applications in all science field. For example, the interferometer is used to study the wave-particle duality of photons [3-5] and the nonclassical effects of quantum sources [6-8]. The fundamental understand of photon interference has long been debated since Dirac [9]. In Dirac's view, photon can only interfere with itself. Such a viewpoint encounters some problems when one explains a two-photon interference generated from a type-I or a type-II spontaneous parametric down conversion process [10, 11]. Later on, physicists updated Dirac's view point as a pair of photon only interferes with the pair itself [11, 12]. Once we known how a photon behaves in a certain interference process, we can better apply this behavior for high precision metrology based on photon interference. Measurements of most physical quantities, including position, displacement, distance, angle, optical dispersion and optical path length, often depend on decoding parameters from a specific interference fringes or patterns [13-18]. How to obtain stable interference fringes and obtain more parameters in a single interference fringe is the long pursued aim in interference based precision optical metrology. For instances, in gravitational-wave detectors like the Laser Interferometer Gravitational-Wave Observatory (LIGO) [19], which must be placed in a high-cost vacuum to achieve high stability. Also in a recent work by Shih's group, which reported a turbulence free double slit experiments based on two-photon interferences [12].

In this letter, an intrinsic stable interferometer is used, which is based on a modified Sagnac interferometer. Since light beams in the two arms of the interferometer are slightly tilted and in counter propagation configurations. Both light beams have nearly the same sensitivity to environmental turbulences such as temperature fluctuation and vibrations. Therefore the phase changes of the two-arms are canceled and the relative phase between the two-arms can hold for hours. By inserting birefringent crystals in this intrinsic stable interferometer, the behaviors of photon interference are studied and revealed for the first time. A complete quantum description of a birefringent MZI is performed in this letter, which is rather different from previous studies [20]. Some interesting results are revealed behind the general theory. A quantum beating versus crystal temperature for both single-photon and two-photon interferences is observed. In addition, the beating intensity can be tuned by rotating the crystal, and the two-photon interference fringes beat two times faster than the single-photon interference fringes. This beating effect is used to determine the thermal dispersion coefficients of the two principal refractive axes with a single measurement. Moreover, the two-photon input case shows super-resolution and higher sensitivity; we also find that the beating behavior is strongly depended on the bandwidth of the input photons, which is come from polarization decoherence of the photons propagation in the birefringent crystal; these behaviors of photon interference have not ever been observed before.

The general theoretical models will be given first. A simplified diagram is shown in Fig. 1, in which two birefringent potassium titanyl phosphate (KTP) crystals have been inserted into a modified MZI. One of these crystals, designated KTP2, is used to compensate for the optical path length differences in the interferometer. The other KTP crystal (KTP1) is mounted on a rotation stage to rotate its position with respect to the horizontal polarization direction. The photon operators at the output ports (P4, P5) are connected to those at the input ports (P0, P1) through the transformations provided by the beam splitters and the birefringent crystals. By assuming that the photon's coherence length is much greater than both the imbalance of the interferometer and the optical path length difference between the optical axes of the birefringent crystal, the photon operators ($\hat{A}_4, \hat{A}_5$) at ports P4 and P5 can be expressed in terms of the operators ($\hat{A}_0, \hat{A}_1$) at ports P0 and P1 using the following expression [21] (for details, see supplementary material [22]):

$$\hat{A}_4 = \frac{1}{\sqrt{2}}(\hat{A}'_2 + i\hat{A}'_3) = F_1(\hat{A}_0, \hat{A}_1)\vec{e}_H + G_1(\hat{A}_0, \hat{A}_1)\vec{e}_V$$
$$\hat{A}_5 = \frac{1}{\sqrt{2}}(i\hat{A}'_2 + \hat{A}'_3) = F_2(\hat{A}_0, \hat{A}_1)\vec{e}_H + G_2(\hat{A}_0, \hat{A}_1)\vec{e}_V \quad (1)$$

where $\hat{F}_1, \hat{G}_1, \hat{F}_2, \hat{G}_2$ are expressed as follows:

$$\hat{F}_1 = -\frac{1}{2}(T-\alpha)\hat{A}_0 + \frac{i}{2}(T+\alpha)\hat{A}_1; \quad \hat{G}_1 = \frac{\beta}{2}(\hat{A}_0 + i\hat{A}_1)$$
$$\hat{F}_2 = \frac{i}{2}(T+\alpha)\hat{A}_0 + \frac{1}{2}(T-\alpha)\hat{A}_1; \quad \hat{G}_2 = \frac{i\beta}{2}(\hat{A}_0 + i\hat{A}_1) \quad (2)$$

where $|\alpha|^2 + |\beta|^2 = 1$, $\alpha = \cos^2\delta e^{i\Delta\varphi_y} + \sin^2\delta e^{i\Delta\varphi_z}$, $\beta = \cos\delta\sin\delta(e^{i\Delta\varphi_y} - e^{i\Delta\varphi_z})$, $\delta$ is the rotation angle of KTP1; $\vec{e}_H$ and $\vec{e}_V$ represent the unit vectors in the two orthogonal polarization directions; and $\Delta\varphi_i = 2\pi n_i(\lambda, T)L/\lambda$ ($i = y, z$) are the optical phase changes along the y and z axes of the birefringent crystal, where L is the crystal length, $\lambda$ is the wavelength of the photon and T is the crystal temperature.

For the single-photon input case, the single-photon count rates $R_4$, $R_5$ at ports 4 and 5 can be calculated to be

$$R_4 = \frac{1}{2}[1 + \cos^2\delta\cos(\Delta\varphi_y - \Delta\varphi_c) + \sin^2\delta\cos(\Delta\varphi_z - \Delta\varphi_c)] \quad (3)$$

$$R_5 = \frac{1}{2}[1 - \cos^2\delta\cos(\Delta\varphi_y - \Delta\varphi_c) - \sin^2\delta\cos(\Delta\varphi_z - \Delta\varphi_c)] \quad (4)$$

For the two-photon input case, the corresponding count rate $R_{4,5}$ between ports 4 and 5 can be calculated as follows:

$$R_{4,5} = \frac{1}{4}\{1 + \cos^4\delta\cos(2\Delta\varphi_y - 2\Delta\varphi_c) + \sin^4\delta[\cos(2\Delta\varphi_z - 2\Delta\varphi_c) + \frac{1}{2}\cos(\Delta\varphi_y + \Delta\varphi_z - 2\Delta\varphi_c)]\} \quad (5)$$

Eqs. (3), (4), and (5) are the main results for input of the narrow-band photon pair. The results for input of a broad-band photon pair will be discussed in the following text and the supplementary material [22].

We now experimentally demonstrate the predictions that were described in the theoretical models above. The experimental setup is shown in Fig. 1. The photon pair is generated using a type-II periodically-poled KTP crystal (PPKTP), which has a length of 2 cm. The crystal has a poling period of 46.2 μm, and the degenerate phase matching temperature for the 775 nm pump beam is 30.030(±0.002)°C. The orthogonally-polarized photons in a pair is separated by the polarizing beam splitter (PBS) and the photons are then coupled into single-mode fibers (SMFs). The pump beam is removed using long pass filters (LPFs) before the photons are coupled into the SMFs. The polarizations of the photons in the SMFs are controlled using two pairs of half-wave plates (HWPs) and quarter-wave plates (QWPs). The photons that are released from the two SMFs are first polarization-purified using two PBSs, and are then injected into a self-stable MZI, which contains two KTP crystals; one KTP crystal is used for the measurements, while the other compensates the optical path length differences between the two arms of the MZI. The self-stable MZI is based on a tilted Sagnac loop, where the clockwise and counterclockwise beams have a traverse distance of 10 mm. The two KTP crystals have dimensions of 5 mm×5 mm×8 mm, and both end faces are anti-reflection coated for 1550 nm. Both crystals are x-cut such that the beams propagate along the x-axes of the crystal. KTP1 is used for the measurements, while KTP2 is used for compensation, and the temperature of KTP2 is kept at a constant 22.300(±0.002)°C. The temperature of KTP1 can be tuned from 17.810(±0.002)°C to 45.670(±0.002)°C. The temperatures of the two crystals are controlled using two home-made temperature controllers with a temperature stability of ±0.002 °C. The delay between the two-photon pairs is controlled using a one-dimensional translation stage. The output photons at ports 4 and 5 are connected to two free-running InGaAs single-photon detectors (SPDs; ID220; 20% quantum efficiency; 5 µs dead time). The output signals from the two SPDs are sent to a coincidence measurement device (Timeharp 260; 0.4 ns coincidence window).

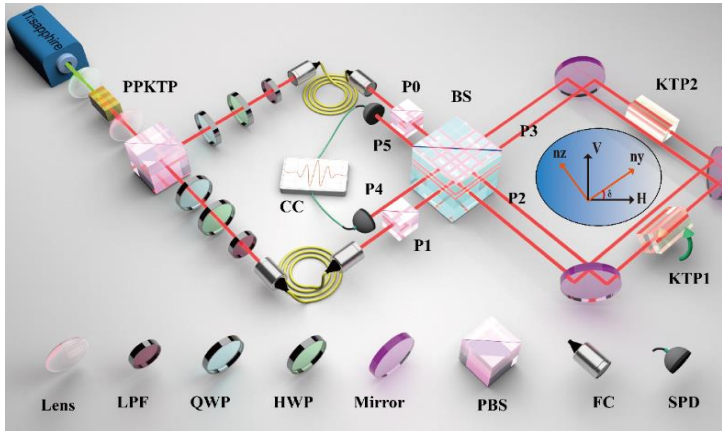

Figure 1. Experimental setup for photon pair generation and photon interference. PPKTP: periodically poled potassium titanyl phosphate crystal; PBS: polarizing beam splitter; SMF: single-mode fibers; FC: fiber collimators; QWP: quarter-wave plates; HWP: half-wave plates LPF: long pass filters; SPD: single-photon detectors.

We firstly characterize the photonic beating effects for the heralded single-photon and two-photon $N00N$ states. The bandwidth of the photon is 1.3 nm without spectral filtering. The Hong-Ou-Mandel (HOM) interference fringes for the photons without filtering and with a 0.5 nm filter are shown in Fig. 2, and we see that nearly perfect HOM interference characteristics are observed in both cases, with visibilities of 97.42±0.16% and 98.71±0.29%, respectively. For HOM measurements, paths P2 and P3 are directly coupled to two SMFs for detection, the delays between the photons are tuned by a one dimensional stage mounted on the signal arm. The HOM dip shapes are determined by the spectrum of the photon pair [10, 23]: for unfiltered case, the spectrum of the photon should be a Sinc$^2$ function [See Eq. (s11) of supplementary material [22]], while for filtered case, the two-photon spectrum is a Gaussian function. In the unfiltered case, the pump power is 8.4 mW, the single count rates are approximately 62 kHz and 77 kHz for the signal and idler photons, respectively, and the dark count rate is approximately 3 kHz. For the filtered case, the pump power is approximately 16 mW, and the single count rates for the signal and idler photons are 90 kHz and 56 kHz, respectively. For the detailed characteristics of the photon sources, please refer to Refs. [24, 25].

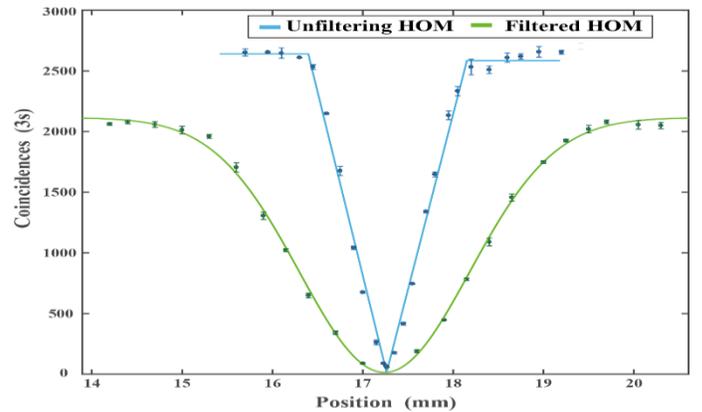

Figure 2. HOM interference fringes for the unfiltered and filtered cases. In the unfiltered case, the theoretical curve used to fit the data is a triangle function; in the filtered case, the theoretical curve used to fit the data is Gaussian.

When the photons in each pair reach the BS simultaneously, a two-photon $N00N$ state is generated after the two output ports of the BS because of bunching effects in HOM interference. The single-photon and two-photon beating curves versus temperature for the different rotation angles of the KTP1 crystal are shown in Fig. 3. The group of figures on the left (bottom to top) shows the beating curves of the two-photon input for rotation angles of $\delta = 0,\ \pi/6,\ \pi/4,\ \pi/3,\ \pi/2$. The group of figures on the right shows the corresponding beating curves for the heralded single photon. The rotation angles of $\delta = 0,\ \pi/2$ represent cases in which the input photon polarization coincides with the *y* and *z* optical axes of the birefringent crystal. The two-photon and single-photon visibilities for the two cases are (98.01±0.18)% and (94.75±0.27)%, and (93.09±0.23)% and (90.87±0.34)%, respectively. In these two cases, the two (single)-photon cases yield thermal dispersions ($dn_y/dT, dn_z/dT$) of (1.027±0.019)×10$^{-5}$/K [(1.041±0.044)×10$^{-5}$/K] and (1.680±0.019)×10$^{-5}$/K [(1.651±0.035)×10$^{-5}$/K] for the *y* and *z* axes, respectively. The other curves show the beating behavior of the optical properties along the two axes, and we can determine the optical properties along both axes from any single measurement of this type of beating curve. For example, when $\delta = \pi/3$, the thermal dispersions that were obtained for two (single) photons for axes *y* and *z* were (0.980±0.067)×10$^{-5}$/K [(0.928±0.104)×10$^{-5}$/K] and (1.592±0.013)×10$^{-5}$/K [(1.594±0.037)×10$^{-5}$/K], respectively.

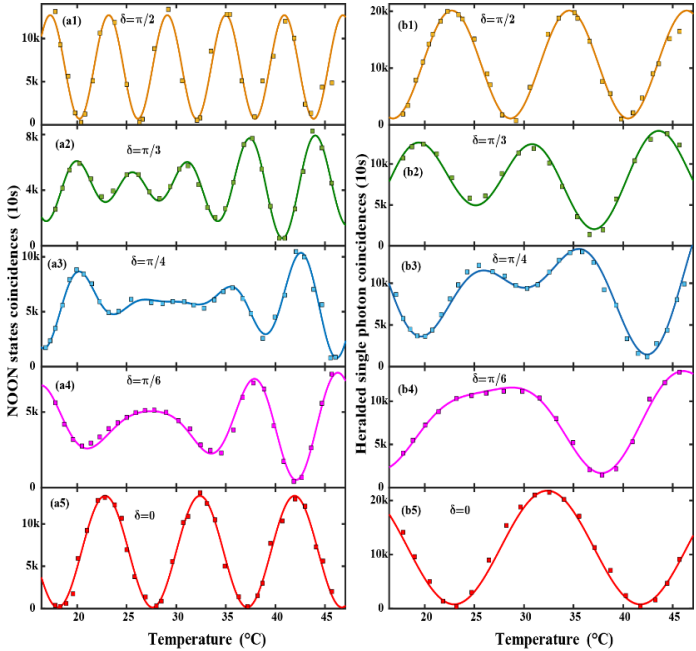

Figure 3. Photonic beating versus temperature for the two-photon input case and the heralded single-photon input case. The group of figures on the left (from bottom to top) represents the two-photon cases at rotation angles of $\delta = 0,\ \pi/6,\ \pi/4,\ \pi/3,\ \pi/2$. The group of figures on the right represents the corresponding heralded single-photon input case. The measurement time is 10s. Different offsets in each of the interference fringes come from the different initial phases between the two arms of the interferometer.

Another important feature of the beating curve is that the temperature oscillation period of the beating curve for the two-photon case is two times faster than that of the heralded single-photon cases, which indicates higher measurement resolution when a high-photon-number entanglement state is used in measurements. We should point out that the environment temperature fluctuation is higher than±0.002 °C (usually ±0.5 °C), but the relative phase of the interferometer keeps unchanged during the measurement time durations, which indicates that the self-stable MZI is immune to environmental turbulences.

Next, we discuss effects of the imbalance of the MZI on the interference fringes for both the two-photon input and the heralded single-photon input when the rotation angle of KTP1 is $\delta = 0$. Three cases are studied: (I) when the optical lengths are equal, the crystal KTP2 is also aligned along the *y* axes, then the optical path difference between two interfering arms is 0 mm; (II) when the imbalance of the MZI is within the single-photon coherence length, which is realized by rotating KTP2 by 90°, and the optical path difference is 0.66 mm; (III) when the imbalance of the MZI is greater than the coherence length of the single photon, which is realized by removing the crystal KTP2, and the optical path length difference is 5.87 mm. In these three cases, the two-photon interference fringes for both the two-photon and single-photon cases are shown in Fig. 4. The visibilities for these three cases for the two-photon and single-photon inputs are 97.98±0.19%, 94.26±0.46%, and 98.18±0.14%, and 93.09±0.23%, 54.20±0.97%, and 0, respectively. Obvious differences between the two-photon interference fringes and the single-photon interference fringes are reveals. The single-photon interference visibility decreases with increasing of optical path difference in the MZI, while the two-photon interference visibility is immune to small optical path differences. As mentioned in the introduction part, single photon is the interference with itself [9] and two-photon interference is the interference of the photon pair itself [11, 12]. The two-photon coherence length is the same as the coherence length of the pump laser beam, which is at the ~km level. Therefore, small optical path differences in the MZI have no effect on the two-photon interference fringes. In contrast, since the single-photon coherence length is at the 2 mm level, the interference fringes are very sensitive to optical path length differences in the MZI.

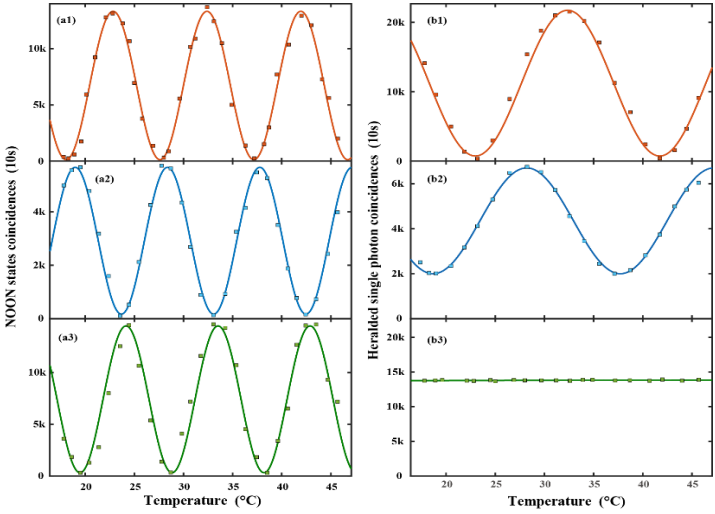

Figure 4. Interference fringes of the two-photon (left) and single-photon (right) cases for $\delta=0$ with increasing optical path length difference. The optical path length differences in order from top to bottom are 0 mm, 0.66 mm and 5. 87 mm.

Finally, we study the effects of the photon bandwidth on the interference fringes, the polarization decoherence of the photon is depended on the photon bandwidth. For photons with an angular spectral bandwidth of $\Delta\omega = 2\sqrt{\ln 2}\sigma$ and a spectral distribution of $f^2(\omega_i) = \frac{1}{\sqrt{\pi}\sigma}\exp(-(\omega_i-\omega_{i0})^2/\sigma^2)$, the single count at port 4 and the coincidence count between port 4 and 5 are given as follows:

$$R_4 = \frac{1}{2}\left(1+\cos^2\delta\cos(\frac{dk_y}{dT}L\Delta T)+\sin^2\delta\exp(-\frac{D^2L^2\sigma^2}{4})\cos(\Delta\varphi+\frac{dk_z}{dT}L\Delta T)\right) \quad (6)$$

$$R_{45} = \frac{1}{2}\{1+\cos^4\delta\cos(2\frac{dk_y}{dT}L\Delta T)+\sin^4\delta\cos(2\frac{dk_z}{dT}L\Delta T+2\Delta\varphi)+ \\ \frac{1}{2}\sin^4\delta\cos[(\frac{dk_z}{dT}+\frac{dk_y}{dT})L\Delta T+\Delta\varphi]\exp(-\frac{D^2L^2\sigma^2}{4})\} \quad (7)$$

where $\Delta\varphi = [k_z(\omega_0,T_0)-k_y(\omega_0,T_0)]L$, $D = \frac{\partial k_z}{\partial\omega}-\frac{\partial k_y}{\partial\omega} = \frac{1}{v_{gz}}-\frac{1}{v_{gy}}$,

and $\Delta T = T-T_0$. These parameters can be obtained from the Sellmeier equations for the $y$ and $z$ axes [26, 27]. The detailed derivations of Eqs. (6) and (7) can be found in the supplementary material [22] and in Ref. [28]. For a photon spectrum with the form of a Sinc$^2$ function, analytical expressions are given by Eqs. (s29)–(s34) in the supplementary material [22]. The experimental results for the unfiltered case for both the two-photon and single-photon input cases when $\delta=\pi/4$ are shown in Fig. 5. The interference visibility decreases as the bandwidth increases. When photon bandwidth is increasing, the time overlapping of photon projected in the two optical axes is decreasing. For photons with large bandwidths, the last terms in Eqs. (6) and (7) can be ignored.

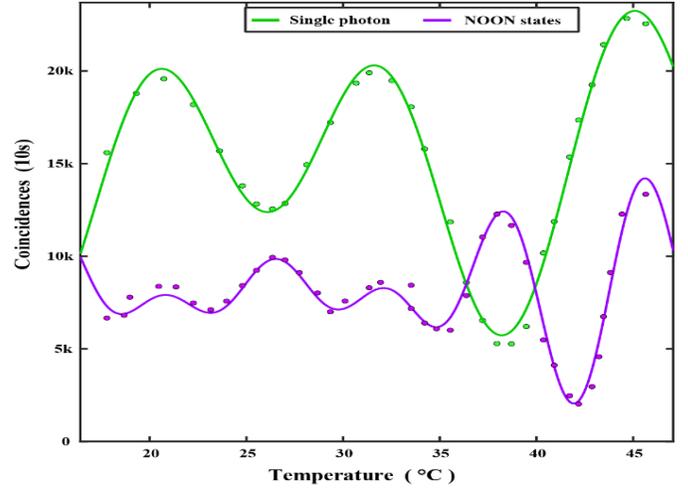

Figure 5. Two-photon and single-photon interferences without filtering when $\delta = \pi/4$. The coincidence time is 10 s. The parameters used for the fitting are L=8 mm, $dn_y/dT, dn_z/dT$ are 1.03×10$^{-5}$/K and 1.620×10$^{-5}$/K, respectively; $\sigma = \frac{\pi}{\sqrt{\ln 2}}\times 50\text{GHz}$; $D$=0.947 ps/mm.

A full theoretical and experimental description of photon interference in a birefringent interferometer is presented in this letter. With careful design of a self-stable MZI, how a photon behaves in the MZI is revealed in details. Photonic beating behavior versus crystal temperature is observed for both single and two-photon. This beating behavior is used to determine the optical properties along both crystal axes with a single measurement. The two-photon interference fringes oscillate twice as fast as those of a single photon, which indicates super-resolution measurement capabilities in the multi-photon entangled state. The differences in single and two-photon interference behaviors in an un-balanced MZI verify the view point that single photon is the interference with itself and two-photon interference is the interference of the photon pair itself. We should point out that the single photon input case is not limited to single photon, the same would be true for lasers, which would simplify the setups in practical applications. The present system is not limited to determination of the thermal dispersion of a birefringent crystal and can also be used to determine the wavelength dispersion [29] and electro-optical coefficient of the birefringent crystal. This study will thus be of great importance for understanding the nature of photon and for precision optical metrology.

**Acknowledgments**

This work is supported by the National Natural Science Foundation of China (NSFC) (61435011, 61525504, 61605194); the National Key Research and Development Program of China (2016YFA0302600); Anhui Initiative In Quantum Information Technologies (AHY020200); the China Postdoctoral Science Foundation (2016M590570); and the Fundamental Research Funds for the Central Universities. We thank David MacDonald, MSc, from Liwen Bianji, Edanz Editing China (www.liwenbianji.cn/ac), for editing the English text of a draft of this manuscript.

# Supplementary of material

**Derivation of single count and coincidence counts for narrow bandwidth photon input**

For a narrow bandwidth photon pair input, the transformation of the annihilation operators when the photon interact at the beam splitter for the first time can be expressed as [21]

$$\hat{A}_2 = \frac{1}{\sqrt{2}}(\hat{A}_0 + i\hat{A}_1)$$
$$\hat{A}_3 = \frac{1}{\sqrt{2}}(i\hat{A}_0 + \hat{A}_1)$$
(s1)

The transformation of birefringence crystal can be expressed as

$$\hat{A}_2' = \hat{A}_2[(\cos^2\delta e^{i\Delta\varphi_y} + \sin^2\delta e^{i\Delta\varphi_z})\vec{e}_H + \cos\delta\sin\delta(e^{i\Delta\varphi_y} - e^{i\Delta\varphi_z})\vec{e}_V]$$
$$= \hat{A}_2[\alpha\vec{e}_H + \beta\vec{e}_V]$$
(s2)

where $|\alpha|^2 + |\beta|^2 = 1$, $\vec{e}_H$ and $\vec{e}_V$ represent the unit vector in the two orthogonal polarization directions; $\Delta\varphi_i = 2\pi n_i(\lambda, T)L/\lambda$ $(i = y, z)$ are the optical phase changes along the y and z axes of the birefringence crystal, where L is the crystal length, $\lambda$ is the wavelength of the photon and T is the temperature of the crystal. The transformation of compensate crystal can be expressed as

$$A_3' = e^{ik(\omega)L_c}A_3 = e^{i\Delta\varphi_c}A_3 = T(\omega)A_3 \quad \text{(s3)}$$

When photons encounter at the BS again, the transformation of the operators can be expressed as

$$\hat{A}_4 = \frac{1}{\sqrt{2}}(\hat{A}_2' + i\hat{A}_3')$$
$$\hat{A}_5 = \frac{1}{\sqrt{2}}(i\hat{A}_2' + \hat{A}_3')$$
(s4)

By using Eqs.(s1-s3), Eq. (s4) can be reduced as

$$\hat{A}_4 = F_1(\hat{A}_0, \hat{A}_1)\vec{e}_H + G_1(\hat{A}_0, \hat{A}_1)\vec{e}_V$$
$$\hat{A}_5 = F_2(\hat{A}_0, \hat{A}_1)\vec{e}_H + G_2(\hat{A}_0, \hat{A}_1)\vec{e}_V$$
(s5)

Where $\hat{F}_1, \hat{G}_1, \hat{F}_2, \hat{G}_2$ are as following:

$$\hat{F}_1 = -\frac{1}{2}(T-\alpha)\hat{A}_0 + \frac{i}{2}(T+\alpha)\hat{A}_1$$
$$\hat{G}_1 = \frac{\beta}{2}(\hat{A}_0 + i\hat{A}_1)$$
$$\hat{F}_2 = \frac{i}{2}(T+\alpha)\hat{A}_0 + \frac{1}{2}(T-\alpha)\hat{A}_1$$
$$\hat{G}_2 = \frac{i\beta}{2}(\hat{A}_0 + i\hat{A}_1)$$
(s6)

Based on Eqs. (s5) and s(6), single photon counts at port 4 and 5 can be expressed as

$$R_4 = \langle 0,1|\hat{A}_4^+\hat{A}_4|0,1\rangle = \langle 0,1|\hat{F}_1^+\hat{F}_1|0,1\rangle + \langle 0,1|\hat{G}_1^+\hat{G}_1|0,1\rangle$$
$$= \frac{1}{2}[1+\cos^2\delta\cos(\Delta\varphi_y-\Delta\varphi_c)+\sin^2\delta\cos(\Delta\varphi_z-\Delta\varphi_c)] \quad (s7)$$

$$R_5 = \langle 0,1|\hat{A}_5^+\hat{A}_5|0,1\rangle = \langle 0,1|\hat{F}_2^+\hat{F}_2|0,1\rangle + \langle 0,1|\hat{G}_2^+\hat{G}_2|0,1\rangle$$
$$= \frac{1}{2}[1-\cos^2\delta\cos(\Delta\varphi_y-\Delta\varphi_c)-\sin^2\delta\cos(\Delta\varphi_z-\Delta\varphi_c)] \quad (s8)$$

Coincidence counts at port 4 and 5,

$$R_{4,5} = \langle 1,1|\hat{A}_4^+\hat{A}_5^+\hat{A}_5\hat{A}_4|1,1\rangle = \langle 1,1|(\hat{F}_1^+\hat{F}_2^+\hat{F}_2\hat{F}_1 + \hat{G}_1^+\hat{G}_2^+\hat{G}_2\hat{G}_1 + \hat{G}_1^+\hat{F}_2^+\hat{F}_2\hat{G}_1 + \hat{F}_1^+\hat{G}_2^+\hat{G}_2\hat{F}_1)|1,1\rangle$$
$$= \frac{1}{2}[1+\cos^4\delta\cos(2\Delta\varphi_y-2\Delta\varphi_c)+\sin^4\delta\cos(2\Delta\varphi_z-2\Delta\varphi_c)+2\sin^2\delta\cos^2\delta\cos(\Delta\varphi_y+\Delta\varphi_z-2\Delta\varphi_c)] \quad (s9)$$

At this point, we obtains all the equations as show in the main text for narrow bandwidth photon input for both single photon and two-photon cases.

**Derivation of single count and coincidence counts for broad bandwidth photon input**

The two photon wave function can be expressed as [28]

$$|\Phi^{(2)}(t)\rangle = \iint d\omega_s d\omega_i \delta(\omega_{0p}-\omega_s-\omega_i) f(\omega_s,\omega_i) e^{i(\omega_s+\omega_i)t} \hat{a}_s^\dagger(\omega_s)\hat{a}_i^\dagger(\omega_i)|0,0\rangle \quad (s10)$$

Where $f(\omega_s,\omega_i)$ is determined by the phase matching condition, and is expressed as

$$f(\omega_s,\omega_i) = \sin c(\frac{\Delta kL}{2}) \quad (s11)$$

The phase mismatching $\Delta k = k_p(\omega_s+\omega_i) - k_s(\omega_s) - k_i(\omega_i) + \frac{2\pi}{\Lambda}$. When heralded single photon is used, the spectral distribution of signal and idler can be expressed as

$$f_s(\omega_s) = \int d\omega_i \delta(\omega_{0p}-\omega_s-\omega_i) f(\omega_s,\omega_i) = f(\omega_s, \omega_{0p}-\omega_s),$$
$$f_i(\omega_i) = \int d\omega_s \delta(\omega_{0p}-\omega_s-\omega_i) f(\omega_s,\omega_i) = f(\omega_i, \omega_{0p}-\omega_i) \quad (s12)$$

Therefore $|\Phi(t)\rangle_i = \int d\omega_i f(\omega_i) e^{i\omega_i t} \hat{a}_i^+ f(\omega_i)|0\rangle$

$$\hat{A}_4(t+t_0) = \int d\omega_i [F_1(\omega_i)\vec{e}_H + G_1(\omega_i)\vec{e}_V] e^{-i\omega_i(t+t_0)} \hat{a}_i(\omega_i)$$

$$K = \langle 0|\hat{A}_4(t+t_0)|\Phi(t_0)\rangle_i = \langle 0|\iint d\omega_i' d\omega_i [F_1(\omega_i')\vec{e}_H + G_1(\omega_i')\vec{e}_V] e^{-i\omega_i'(t+t_0)} f(\omega_i) e^{i\omega_i t_0} \hat{a}_i(\omega_i')\hat{a}_i^+(\omega_i)|0\rangle$$
$$= \int d\omega_i [F_1(\omega_i)\vec{e}_H + G_1(\omega_i)\vec{e}_V] f(\omega_i) e^{-i\omega_i t} \quad (s13)$$

For single photon input at port 1, the single count at port 4 can be expressed as

$$R_4 = \int dt_i |K|^2 = \int dt_i \left| \int d\omega_i [F_1(\omega_i)\vec{e}_H + G_1(\omega_i)\vec{e}_V] f(\omega_i) e^{-i\omega_i t} \right|^2$$

$$= \frac{1}{4} \int d\omega_i \left( |T(\omega_i) + \alpha(\omega_i)|^2 + |\beta(\omega_i)|^2 \right) f^2(\omega_i) \quad \text{(s14)}$$

$$= \frac{1}{2} \int d\omega_i \left( 1 + \cos^2\delta \cos(\Delta\varphi_y - \Delta\varphi_c) + \sin^2\delta \cos(\Delta\varphi_z - \Delta\varphi_c) \right) f^2(\omega_i)$$

When narrow band photon pairs are used, the frequency distribution of the photon pairs can be expressed as $f^2(\omega_i) = \delta(\omega - \omega_{i0})$, and the above expression can be expressed as

$$R_4 = \frac{1}{4}\left( |T(\omega_0) + \alpha(\omega_0)|^2 + |\beta(\omega_0)|^2 \right)$$

$$= \frac{1}{2}\left( 1 + \cos^2\delta \cos(\Delta\varphi_y - \varphi_c) + \sin^2\delta \cos(\Delta\varphi_z - \varphi_c) \right) \quad \text{(s15)}$$

Equation (s16) recovery the case of narrow bandwidth photon input case.

For a single photon with Gaussian spectrum $f^2(\omega_i) = \frac{1}{\sqrt{\pi}\sigma} \exp(-(\omega_i - \omega_{i0})^2/\sigma^2)$, and bandwidth of photon is $\Delta\omega = 2\sqrt{\ln 2}\sigma$. The phase changes are depended on frequency and temperature, therefore we can expanded them at the central frequency and the temperature of the compensate crystal.

$$\Delta\varphi_y = k_y(\omega, T)L = [k_y(\omega_0, T_0) + \frac{\partial k_y}{\partial \omega}(\omega - \omega_0) + \frac{\partial k_y}{\partial T}(T - T_0)]L$$

$$\Delta\varphi_c = k_y(\omega, T_0)L = [k_y(\omega_0, T_0) + \frac{\partial k_y}{\partial \omega}(\omega - \omega_0)]L \quad \text{(s16)}$$

$$\Delta\varphi_z = k_z(\omega, T)L = [k_z(\omega_0, T) + \frac{\partial k_z}{\partial \omega}(\omega - \omega_0) + \frac{\partial k_z}{\partial T}(T - T_0)]L$$

Therefore the single count rate at port 4 for photon with broad bandwidth can be expressed as

$$R_4 = \frac{1}{2}\left( 1 + \cos^2\delta \cos(\frac{dk_y}{dT}L\Delta T) + \sin^2\delta \exp(-\frac{D^2 L^2 \sigma^2}{4}) \cos(\Delta\varphi + \frac{dk_z}{dT}L\Delta T) \right) \quad \text{(s17)}$$

Where $\Delta\varphi = [k_z(\omega_0, T_0) - k_y(\omega_0, T_0)]L$, $D = \frac{\partial k_z}{\partial \omega} - \frac{\partial k_y}{\partial \omega} = \frac{1}{v_{gz}} - \frac{1}{v_{gy}}$, $\Delta T = T - T_0$.

Similar to the derivation of equation (s18), the single count rate at port five can be expressed as

$$\langle \Phi | \hat{A}_5^+ \hat{A}_5 | \Phi \rangle = \int dt_i \left| \int d\omega_i [F_2(\omega_i)\vec{e}_H + G_2(\omega_i)\vec{e}_V] f(\omega_i) e^{-i\omega_i t} \right|^2$$

$$= \frac{1}{4} \int d\omega_i \left( |T(\omega_i) - \alpha(\omega_i)|^2 + |\beta(\omega_i)|^2 \right) f^2(\omega_i) \quad \text{(s18)}$$

$$R_5 = \frac{1}{2}\left( 1 - \cos^2\delta \cos(\frac{dk_y}{dT}L\Delta T) - \sin^2\delta \exp(-\frac{D^2 L^2 \sigma^2}{4}) \cos(\Delta\varphi + \frac{dk_z}{dT}L\Delta T) \right) \quad \text{(s19)}$$

For two-photon input case, the coincidence count at port 4 and 5 can be expressed as

$$\iint dt_s dt_i \langle \Phi(t_s,t_i)| \hat{A}_4^+ \hat{A}_5^+ \hat{A}_5 \hat{A}_4 |\Phi(t_s,t_i)\rangle$$
$$= \iint dt_s dt_i \langle \Phi(t_s,t_i)|(\hat{F}_1^+ \hat{F}_2^+ \hat{F}_2 \hat{F}_1 + \hat{G}_1^+ \hat{G}_2^+ \hat{G}_2 \hat{G}_1 + \hat{G}_1^+ \hat{F}_2^+ \hat{F}_2 \hat{G}_1 + \hat{F}_1^+ \hat{G}_2^+ \hat{G}_2 \hat{F}_1)|\Phi(t_s,t_i)\rangle \quad (s20)$$

Where

$$\iint dt_s dt_i \langle \Phi(t_s,t_i)|\hat{F}_1^+ \hat{F}_2^+ \hat{F}_2 \hat{F}_1 |\Phi(t_s,t_i)\rangle =$$
$$\frac{1}{4}\iint d\omega_s d\omega_i f^2(\omega_s,\omega_i)\delta(\omega_{0p} - \omega_s - \omega_i)|T(\omega_s)T(\omega_i) + \alpha(\omega_s)\alpha(\omega_i)|^2 \quad (s21)$$

$$\iint dt_s dt_i \langle \Phi(t_s,t_i)|\hat{G}_1^+ \hat{G}_2^+ \hat{G}_2 \hat{G}_1 |\Phi(t_s,t_i)\rangle$$
$$= \frac{1}{4}\iint d\omega_s d\omega_i f^2(\omega_s,\omega_i)\delta(\omega_{0p} - \omega_s - \omega_i)|\beta(\omega_s)\beta(\omega_i)|^2 \quad (s22)$$

$$\hat{G}_1^+ \hat{F}_2^+ \hat{F}_2 \hat{G}_1 = \frac{1}{16}|T(\omega_i)\beta(\omega_s) - \alpha(\omega_i)\beta(\omega_s) - T(\omega_s)\beta(\omega_i) - \alpha(\omega_s)\beta(\omega_i)|^2 \quad (s23)$$

$$\hat{F}_1^+ \hat{G}_2^+ \hat{G}_2 \hat{F}_1 = \frac{1}{16}|\beta(\omega_s)T(\omega_i) + \beta(\omega_s)\alpha(\omega_i) - \beta(\omega_i)T(\omega_s) + \alpha(\omega_s)\beta(\omega_i)|^2 \quad (s24)$$

$$\iint dt_s dt_i \langle \Phi(t_s,t_i)|\hat{G}_1^+ \hat{F}_2^+ \hat{F}_2 \hat{G}_1 + \hat{F}_1^+ \hat{G}_2^+ \hat{G}_2 \hat{F}_1 |\Phi(t_s,t_i)\rangle$$
$$= \frac{1}{8}\iint d\omega_s d\omega_i f^2(\omega_s,\omega_i)\delta(\omega_{0p} - \omega_s - \omega_i)(|\beta(\omega_s)T(\omega_i) - T(\omega_s)\beta(\omega_i)|^2 + |\alpha(\omega_i)\beta(\omega_s) + \alpha(\omega_s)\beta(\omega_i)|^2)$$
(s25)

$$\Gamma(\omega_s,\omega_i) = \hat{F}_1^+ \hat{F}_2^+ \hat{F}_2 \hat{F}_1 + \hat{G}_1^+ \hat{G}_2^+ \hat{G}_2 \hat{G}_1 + \hat{G}_1^+ \hat{F}_2^+ \hat{F}_2 \hat{G}_1 + \hat{F}_1^+ \hat{G}_2^+ \hat{G}_2 \hat{F}_1$$
$$= \frac{1}{2}\{1 + \cos^4 \delta \cos(y_s + y_i - y_{s0} - y_{i0}) + \sin^4 \delta \cos(z_s + z_i - y_{s0} - y_{i0}) +$$
$$2\cos^2 \delta \sin^2 \delta \cos(\frac{z_s - z_i - y_s + y_i}{2})\cos(\frac{z_s + z_i + y_s + y_i - 2y_{s0} - 2y_{i0}}{2}) + \quad (s26)$$
$$2\cos^2 \delta \sin^2 \delta \sin(\frac{z_i - y_i}{2})\sin(\frac{z_s - y_s}{2})[\cos(\frac{z_s - z_i - y_s + y_i}{2}) - \cos(\frac{z_s - z_i + y_s - y_i - 2y_{s0} + 2y_{i0}}{2})]\}$$

Where the symbols in Eq. (s26) are as follows:

$$y_j = k_y(\omega_j,T)L = [k_y(\omega_{j0},T_0) + \frac{\partial k_y}{\partial \omega}(\omega_j - \omega_{j0}) + \frac{\partial k_y}{\partial T}(T - T_0)]L$$

$$y_{j0} = k_y(\omega_j,T_0)L = [k_y(\omega_{j0},T_0) + \frac{\partial k_y}{\partial \omega}(\omega_j - \omega_{j0})]L \qquad (j = s,i) \quad (s27)$$

$$z_j = k_z(\omega_j,T)L = [k_z(\omega_{j0},T) + \frac{\partial k_z}{\partial \omega}(\omega_j - \omega_{j0}) + \frac{\partial k_z}{\partial T}(T - T_0)]L$$

After some detail calculation, the coincidence count at port 4 and 5 can be expressed as

$$\iint dt_s dt_i \langle \Phi(t_s,t_i)|\hat{A}_4^+\hat{A}_5^+\hat{A}_5\hat{A}_4|\Phi(t_s,t_i)\rangle$$
$$= \iint d\omega_s d\omega_i f^2(\omega_s,\omega_i)\delta(\omega_{0p}-\omega_s-\omega_i)\Gamma(\omega_s,\omega_i)$$
$$= \frac{1}{2}\{1+\cos^4(\delta)\cos(2\frac{dk_y}{dT}L\Delta T)+\sin^4(\delta)\cos(2\frac{dk_z}{dT}L\Delta T+2\Delta\varphi)+ \quad (s28)$$
$$2\cos^2(\delta)\sin^2(\delta)\cos[(\frac{dk_z}{dT}+\frac{dk_y}{dT})L\Delta T+\Delta\varphi]\exp(-\frac{D^2L^2\sigma^2}{4})\}$$

The above equations (s17), (s19) and (s28) show the influence of bandwidth of photon pairs to single counts and coincidence counts. Eq. (s26) is a general equation to describe the basic properties of a birefringence MZI, this formula can be used to study other optical properties of a birefringence crystal such as wavelength dispersion and electrical-optical effects. The above formula are calculated for photon pairs with Gaussian spectrum. Actually for photon with spectrum of Sinc² function $f^2(\omega_i)=\frac{\gamma}{2\pi}\text{sinc}^2[\gamma(\omega_i-\omega_{i0})]$, where $\gamma=DL_{spdc}/2$ is determined by the parameters of the SPDC crystal. The expressions for single count rate and coincidence count rate is as following for $L_{spdc}>L$:

$$R_4=\frac{1}{2}\left(1+\cos^2\delta\cos(\frac{dk_y}{dT}L\Delta T)+\sin^2\delta\frac{L_{spdc}-L}{L_{spdc}}\cos(\Delta\varphi+\frac{dk_z}{dT}L\Delta T)\right) \quad (s29)$$

$$R_5=\frac{1}{2}\left(1-\cos^2\delta\cos(\frac{dk_y}{dT}L\Delta T)-\sin^2\delta\frac{L_{spdc}-L}{L_{spdc}}\cos(\Delta\varphi+\frac{dk_z}{dT}L\Delta T)\right) \quad (s30)$$

$$R_{45}=\frac{1}{2}\{1+\cos^4(\delta)\cos(2\frac{dk_y}{dT}L\Delta T)+\sin^4(\delta)\cos(2\frac{dk_z}{dT}L\Delta T+2\Delta\varphi)+$$
$$2\cos^2(\delta)\sin^2(\delta)\cos[(\frac{dk_z}{dT}+\frac{dk_y}{dT})L\Delta T+\Delta\varphi]\frac{L_{spdc}-L}{L_{spdc}}\} \quad (s31)$$

While for $L_{spdc}<L$, expressions (s29) to (s31) become

$$R_4=\frac{1}{2}\left(1+\cos^2\delta\cos(\frac{dk_y}{dT}L\Delta T)\right) \quad (s32)$$

$$R_5=\frac{1}{2}\left(1-\cos^2\delta\cos(\frac{dk_y}{dT}L\Delta T)\right) \quad (s33)$$

$$R_{45}=\frac{1}{2}[1+\cos^4(\delta)\cos(2\frac{dk_y}{dT}L\Delta T)+\sin^4(\delta)\cos(2\frac{dk_z}{dT}L\Delta T+2\Delta\varphi)] \quad (s34)$$